\documentclass[11pt]{article}
\usepackage[a4paper,margin=2.4cm]{geometry}
\usepackage{amsmath,amssymb,amsthm}
\usepackage{graphicx}
\usepackage{booktabs}
\usepackage{siunitx}
\usepackage{multirow}
\usepackage{xcolor}
\usepackage{tikz}
\usetikzlibrary{arrows.meta,positioning,fit}
\usepackage[section]{placeins}
\usepackage[colorlinks=true,linkcolor=blue!50!black,citecolor=blue!50!black,urlcolor=blue!50!black]{hyperref}

\newtheorem{theorem}{Theorem}
\newtheorem{proposition}{Proposition}
\newcommand{\dmu}{\,\mathrm{d}\mu}
\newcommand{\norm}[1]{\lVert #1\rVert}

\title{Certificate-Based Verification of Pulsed Field Ablation Models}
\author{Konstantinos Dovelos\\ \small Dovelos Engineering \\ \small \texttt{konstantinos.dovelos@gmail.com}}
\date{}

\begin{document}
\maketitle

\begin{abstract}
Computational models of cardiac pulsed field ablation predict the lesion as the
tissue where the electric field exceeds a lethal threshold. Their numerical
error is rarely bounded, and automated pipelines increasingly run them without
an expert checking each result. To verify each run, we separate the question
``was the stated model solved correctly?'' from the solver that answers it. An
untrusted solver returns its answer with a certificate, which consists of a potential
with the correct electrode and ground values and an exactly charge-conserving
current. A
small independent checker verifies the certificate against a written
specification of the physics. Using the Prager--Synge hypercircle theorem
evaluated in ball arithmetic, it turns the certificate into guaranteed
two-sided bounds on the impedance and on the root-mean-square field of every
tile of a lesion map, also over a range of blood/tissue conductivity ratios. The
bounds cover numerical error only; whether the model describes tissue is a
question of validation. A vectorized version of the checker takes about a tenth of our solver's time. For
a focal electrode in contact, the tile-resolved depth at \SI{1000}{V} and
\SI{700}{V/cm} is proven to lie in \SIrange{3.50}{3.75}{mm} for every ratio from
1.7 to 3.75. For a bipolar catheter, certified depths overlap published in-vivo
depths. Where they differ from a published reference model, the certificates
prove that the difference is not our numerical error. We compare
the certificate with a grid convergence study on the same solver. On graded grids,
the study is sharper for local fields but remains an estimate. On uniform grids, it
fails to converge at the electrode edge, where the certificate still gives a valid
bracket. Solvers that deliberately solve equations other than the specified ones
pass the grid convergence study, but the certificates refute the impedances they
report or leave them unsupported.
\end{abstract}

\noindent\textbf{Keywords:} a posteriori error estimation, cardiac pulsed field
ablation, computational modeling, electroporation, medical device verification,
verified numerics.

\section{Introduction}

Pulsed field ablation (PFA) kills myocardium by irreversible electroporation
where
the electric field exceeds a lethal threshold, so lesion size is predicted by
solving a quasi-static conduction problem and taking a threshold contour
\cite{meckes2022,petras2025}. Finite-element simulation is the standard tool
for mapping the field of a PFA catheter, while the lethal threshold and tissue
response remain the main sources of model uncertainty~\cite{anic2026}.
Such models are used to compare catheter
geometries, to choose voltages, and, increasingly, to support regulatory
submissions. The verification and validation (V\&V) standard V\&V~40 of the
American Society of Mechanical Engineers and the guidance of the U.S.\ Food and
Drug Administration on computational modeling set risk-informed credibility
goals, under which the more a decision relies on a
model, and the graver its consequence, the more evidence is needed
\cite{asmevv40,fda2023}.

On the numerical side, evidence for accuracy comes at two levels
\cite{roache1998,oberkampf2010}. Code verification, for example by manufactured
solutions and order-of-accuracy tests, shows that the software can solve its
equations correctly. Calculation verification estimates the discretization,
solver, and round-off error of the simulations actually used. It is commonly a
mesh refinement study (mesh study for short) on a representative case, with the
result assumed to carry over to the runs in the context of use. It gives
estimates rather than bounds, and repeating it for every run of a design loop is
costly. Guaranteed a posteriori bounds make calculation verification rigorous.
They prove a bound on the error of a particular run, a small program independent
of the solver can check them, and they can be attached to every run rather than
to a representative one. This paper concerns such rigorous calculation verification.

Estimates have been acceptable because an expert chose the mesh and checked each
result. Simulation is increasingly automated, by scripted design loops, by solvers written
with artificial intelligence, and by learned surrogates, i.e., machine-learning
models such as neural operators that predict full fields in seconds
\cite{lu2021}. Automated simulations are not known to be
less accurate, but once no expert checks each result, the per-run check is
lost. This paper does not test solvers written by artificial intelligence or
learned surrogates, which a companion paper does \cite{dovelos2026b}. Here,
solvers with deliberately wrong physics stand in for the former, and surrogates
are discussed in Section~\ref{sec:discussion}.

We take the view of proof-carrying code \cite{necula1997} and of certifying
algorithms \cite{mcconnell2011}. In this view, the solver is untrusted, and its
output carries evidence that a small checker can verify (Section~\ref{sec:cert}). For linear conduction, the
evidence is classical. The hypercircle of Prager and Synge
\cite{pragersynge1947,synge1957} bounds the energy error of any admissible
potential by its distance to any admissible, exactly conserved current. The
complementary Dirichlet and Thomson principles give two-sided bounds on
capacities and impedances \cite{polya1951}.
Equilibrated-flux estimators built on these ideas are mature
\cite{ladeveze1983,braess2009,ernvohralik2015}. Functional majorants extend
them to arbitrary approximations and to problems with uncertain data
\cite{repin2008,mali2014}. Verified computation combines hypercircle bounds with
interval arithmetic to prove statements about partial differential equations
(PDEs)
\cite{liuoishi2013,liu2021stokes,nakao2019}. Certified reduced-basis methods bound outputs
over parameter domains \cite{rozza2008}, and neural PDE solvers have been given
error majorants \cite{fanaskov2024}. Equilibrated estimators have been extended
to parametric problems \cite{eigel2016} and to include the algebraic error of
inexact solves \cite{papez2018}, and they certify existence and outputs for
semilinear problems by post-processing a conventional finite-element solution in
floating point \cite{ishizaka2026}. Checking the output of an untrusted solver by an
independent program is established in exact optimization \cite{cheung2017}.

None of these ingredients is new. Our contribution is their assembly into a
small checker, independent of the solver, that turns any admissible
potential-current pair into guaranteed bounds on the impedance and on the
tile-resolved lesion map of a PFA electrode, including rounding and over
ranges of conductivity. Solvers whose equations deliberately differ from the
specification converge under refinement and pass a grid convergence study, but
the checker refutes the impedance that each solver reports, which we call its
claim, or leaves it unsupported. To the best of our
knowledge, these are the first lesion maps for PFA with proven numerical error
bounds. In particular, the paper provides:
\begin{itemize}
\item A written specification of the physics, kept separate from every
numerical choice, against which certificates are checked.
\item Exact integration in ball arithmetic \cite{arb}, so the bounds are proven
rather than estimated, including rounding.
\item A vectorized checker in double-precision ball arithmetic, rigorous under
stated assumptions, that checks the lesion map in about a tenth of our solver's time
and a range certificate in about a fifth of the time of its solves.
\item Certified lesion maps (proven lethal, proven spared, undecided), stated for
the root-mean-square field of each tile rather than pointwise
(Section~\ref{sec:model}).
\item Bounds that hold over a range of tissue conductivities, obtained by blending certificates.
\item A comparison with a standard grid convergence study on the same solver,
including solvers with deliberately wrong physics that the study accepts and
the certificates do not.
\item A first independence check, in which the unchanged checker certifies a potential
computed by FEniCSx on the same grid, with bounds identical to those for our
solver (Table~\ref{tab:fenics}).
\end{itemize}

Certificates bound the numerical error of the stated model. Whether the model
and threshold describe tissue is a separate question of validation, discussed in
Section~\ref{sec:discussion}.

\section{Model and Lesion Criterion}\label{sec:model}

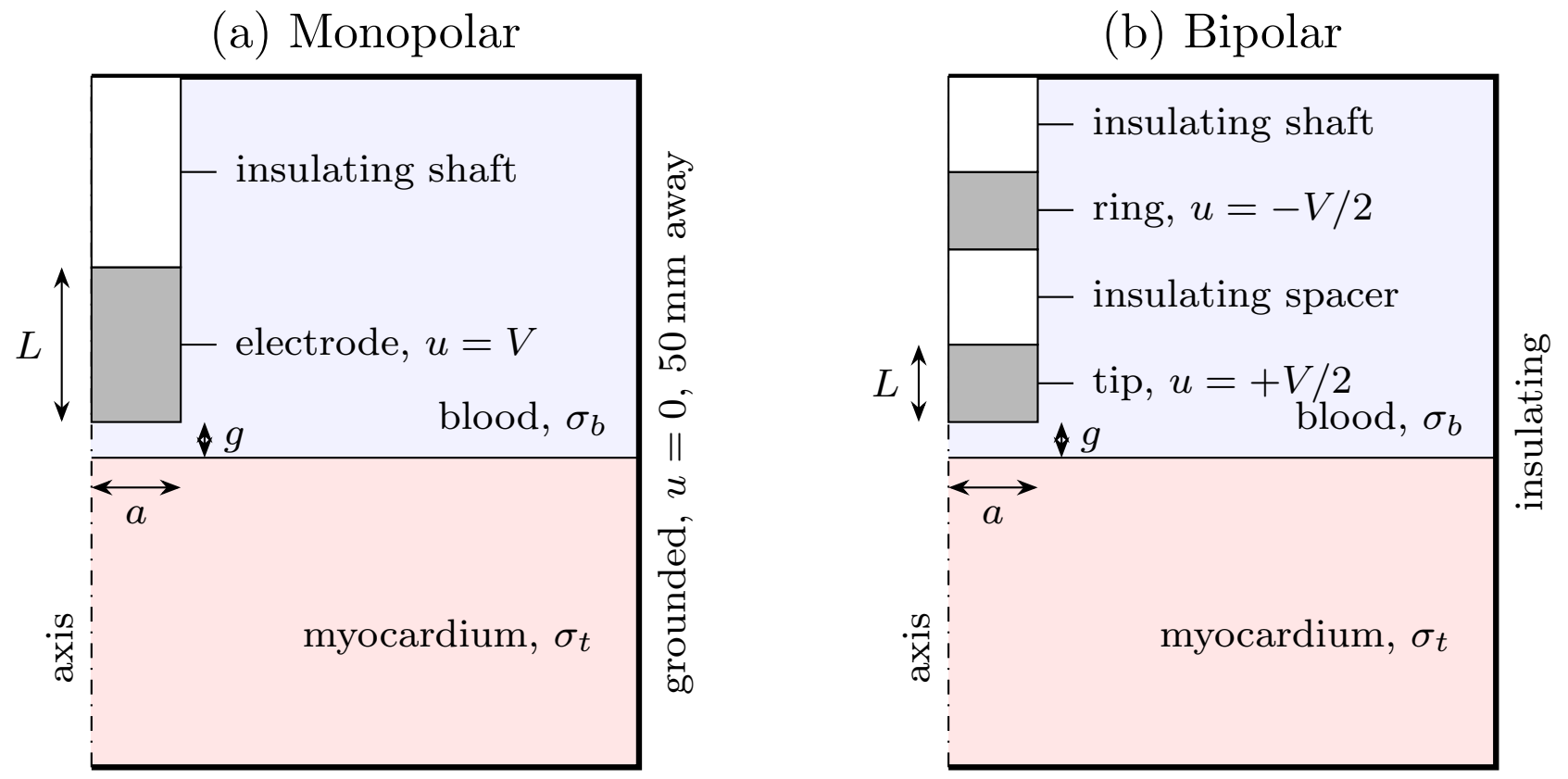
\begin{figure}[!t]
\centering
\begin{tikzpicture}[scale=0.85, font=\scriptsize, >={Stealth[length=1.5mm]}]
\begin{scope}[xshift=0cm]
\fill[red!10] (0,-2.6) rectangle (4.6,0);
\fill[blue!5] (0,0) rectangle (4.6,3.2);
\draw[thin] (0,0) -- (4.6,0);
\draw[very thick] (0,3.2) -- (4.6,3.2) -- (4.6,-2.6) -- (0,-2.6);
\draw[dash dot] (0,-2.6) -- (0,3.2);
\filldraw[fill=gray!55] (0,0.3) rectangle (0.75,1.6);
\draw (0.75,0.95) -- (1.05,0.95) node[right] {electrode, $u=V$};
\filldraw[fill=white] (0,1.6) rectangle (0.75,3.2);
\draw (0.75,2.4) -- (1.05,2.4) node[right] {insulating shaft};
\draw[<->] (-0.25,0.3) -- (-0.25,1.6) node[midway, left] {$L$};
\draw[<->] (0.95,0) -- (0.95,0.3) node[midway, right] {$g$};
\draw[<->] (0,-0.25) -- (0.75,-0.25) node[midway, below] {$a$};
\node[rotate=90, anchor=south] at (0,-1.6) {axis};
\node at (3.0,-1.5) {myocardium, $\sigma_t$};
\node[anchor=south east] at (4.5,0.02) {blood, $\sigma_b$};
\node[rotate=90, anchor=north] at (4.6,0.3) {grounded, $u=0$, \SI{50}{mm} away};
\node[font=\small] at (2.3,3.55) {(a) Monopolar};
\end{scope}
\begin{scope}[xshift=7.2cm]
\fill[red!10] (0,-2.6) rectangle (4.6,0);
\fill[blue!5] (0,0) rectangle (4.6,3.2);
\draw[thin] (0,0) -- (4.6,0);
\draw[very thick] (0,3.2) -- (4.6,3.2) -- (4.6,-2.6) -- (0,-2.6);
\draw[dash dot] (0,-2.6) -- (0,3.2);
\filldraw[fill=gray!55] (0,0.3) rectangle (0.75,0.95);
\draw (0.75,0.625) -- (1.05,0.625) node[right] {tip, $u=+V/2$};
\filldraw[fill=white] (0,0.95) rectangle (0.75,1.75);
\draw (0.75,1.35) -- (1.05,1.35) node[right] {insulating spacer};
\filldraw[fill=gray!55] (0,1.75) rectangle (0.75,2.4);
\draw (0.75,2.08) -- (1.05,2.08) node[right] {ring, $u=-V/2$};
\filldraw[fill=white] (0,2.4) rectangle (0.75,3.2);
\draw (0.75,2.8) -- (1.05,2.8) node[right] {insulating shaft};
\draw[<->] (-0.25,0.3) -- (-0.25,0.95) node[midway, left] {$L$};
\draw[<->] (0.95,0) -- (0.95,0.3) node[midway, right] {$g$};
\draw[<->] (0,-0.25) -- (0.75,-0.25) node[midway, below] {$a$};
\node[rotate=90, anchor=south] at (0,-1.6) {axis};
\node at (3.0,-1.5) {myocardium, $\sigma_t$};
\node[anchor=south east] at (4.5,0.02) {blood, $\sigma_b$};
\node[rotate=90, anchor=north] at (4.6,0.3) {insulating};
\node[font=\small] at (2.3,3.55) {(b) Bipolar};
\end{scope}
\end{tikzpicture}
\caption{Axisymmetric model in the $(r,z)$ half-plane, not to scale. (a) Monopolar:
the electrode is held at $V$ and the outer boundary, \SI{50}{mm} away, is
grounded. (b) Bipolar: tip and ring electrodes at $\pm V/2$ are separated by an
insulating spacer, and the outer boundary is insulating. The axis is a symmetry
line, and the electrode sits a gap $g\ge0$ above the tissue.}
\label{fig:geometry}
\end{figure}

\subsection{Geometry and Equations}

We use cylindrical coordinates $(r,z)$ with rotational symmetry and the measure
$\mathrm{d}\mu = 2\pi r\,\mathrm{d}r\,\mathrm{d}z$. Myocardium (conductivity
$\sigma_t$) fills $z<0$ and blood ($\sigma_b$) fills $z>0$. A flat-tipped
cylindrical metal electrode of radius $a=\SI{1.25}{mm}$ and length $L$ sits a
gap $g\ge0$ above the tissue and continues as an insulating shaft. In the
monopolar configuration, the electrode is held at $V$ and the outer boundary,
\SI{50}{mm} away, is grounded. In the bipolar configuration, an insulating
spacer separates a tip electrode from a ring electrode. The generator floats,
so the outer boundary is insulating and only the potential difference $V$
matters; we hold the tip at $+V/2$ and the ring at $-V/2$ without loss of
generality. Figure~\ref{fig:geometry} shows both configurations. The potential $u$ solves
\begin{equation}
\begin{aligned}
-\nabla\cdot(\sigma\nabla u)&=0 &&\text{in the domain},\\
u&=u_D &&\text{on }\Gamma_D,\\
\sigma\partial_n u&= 0 &&\text{on }\Gamma_N,
\end{aligned}
\end{equation}
with $\Gamma_D$ the electrodes (and ground) and $\Gamma_N$ the insulating walls
and the axis. All boundaries are axis-aligned, so the problem can be discretized on a
rectangular grid. Electrodes are voltage-controlled and conductivities are
constant, so the problem is linear in $V$, and one certificate at \SI{1}{V}
serves all voltages.

\subsection{Lesion Criterion} A lesion-map tile $\omega$ is a ring of square
cross-section, \SI{0.25}{mm} wide, in one material. It is lethal when its
root-mean-square (RMS) field $E_\omega = (\int_\omega |\nabla u|^2\dmu/|\omega|)^{1/2}$, with $|\omega|$ its volume, reaches the
threshold $E_\mathrm{th}$. Given a certified interval for $E_\omega$, a tile is
proven lethal, proven spared, or undecided as its interval lies above, below, or
across the threshold. We report two depths. The depth
``reached'' is the bottom of the deepest axis tile such that it and every axis
tile above it are proven lethal, so the tile-resolved lesion is at least this deep on
the axis.
The depth ``not beyond'' is the level below which every tile of the map, in every
column, is proven spared; the statement is limited to the mapped region. Lesion radii are defined in the same way along each row of
tiles. The axis tiles are the smallest and carry the widest intervals. The
proven-lethal inner envelope and the proven-spared outer envelope bound the
tile-resolved lesion from inside and outside.

This criterion coarsens the pointwise threshold contour used by Meckes et al.\
and Petras et al.\ \cite{meckes2022,petras2025}. In a proven-lethal tile,
$|\nabla u|\ge E_\mathrm{th}$ on a set of positive measure, and in a proven-spared
tile, $|\nabla u|<E_\mathrm{th}$ on a set of positive measure. Neither statement
holds for every point of the tile, so a proven-spared tile may still contain
points above the threshold, and showing that collateral tissue is spared needs
pointwise bounds (Section~\ref{sec:discussion}). The envelopes locate the
pointwise contour to within about one tile where the field decreases
monotonically along the axis. We use \SI{700}{V/cm} as an
illustrative threshold between the ranges that Petras et al.\ find for depth
(\SIrange{270}{500}{V/cm}) and width (\SIrange{790}{1000}{V/cm}) \cite{petras2025}, and \SI{268}{V/cm} where we
follow Meckes et al.

\section{Certificates and the Checker}\label{sec:cert}

\begin{figure}[t]
\centering
\begin{tikzpicture}[font=\small, node distance=5mm and 6mm,
  box/.style={draw, rounded corners=2pt, align=flush center, inner sep=3pt, minimum height=12mm},
  untrusted/.style={box, fill=orange!10},
  trusted/.style={box, fill=blue!8, thick},
  test/.style={box, fill=blue!8, text width=44mm, minimum height=15mm},
  arr/.style={-{Stealth[length=2mm]}, thick}]
\node[untrusted, text width=22mm] (sol) {\textbf{Solver}\\
  {\scriptsize finite elements of degree $p$, or any other method}};
\node[box, text width=26mm, right=of sol] (cert) {\textbf{Certificate}\\
  {\scriptsize nodal values of $v$ and $\psi$ on each cell; one per conductivity ratio}};
\node[test, right=8mm of cert, yshift=11mm] (t1) {\textbf{1.~Specification check}\\
  {\scriptsize boundary values of $v$ and $\psi$ tested exactly; the current conserves charge by construction}};
\node[test, below=7mm of t1] (t2) {\textbf{2.~Numerical verification}\\
  {\scriptsize hypercircle bounds for the specified equations, in ball arithmetic}};
\node[font=\small\bfseries, above=1mm of t1] (chklabel) {Checker};
\node[draw, dashed, thick, rounded corners=3pt, inner sep=2.5mm, fit=(chklabel)(t1)(t2)] (chk) {};
\node[trusted, text width=64mm, above=5mm of chk] (spec) {\textbf{Specification}\\
  {\scriptsize geometry, conductivities, electrode and ground values}};
\node[box, text width=36mm, anchor=west, fill=red!8] (rej) at ([xshift=7mm]chk.east |- t1) {\textbf{Rejected}\\
  {\scriptsize not admissible, e.g.\ a wrong boundary value; no bounds}};
\node[box, text width=36mm, anchor=west, fill=green!8] (bnd) at ([xshift=7mm]chk.east |- t2) {\textbf{Proven bounds}\\
  {\scriptsize impedance interval; RMS-field interval of every tile}};
\node[box, text width=36mm, below=4mm of bnd] (claims) {\textbf{Claims judged}\\
  {\scriptsize a solver's impedance: certified, refuted, or unsupported; each tile: proven lethal, proven spared, or undecided}};
\draw[arr] (sol) -- (cert);
\draw[arr] (cert.east) -- (cert.east -| chk.west);
\draw[arr] (spec) -- (chk);
\draw[arr] (t1) -- node[right, font=\scriptsize] {passes} (t2);
\draw[arr] (t1.east) -- node[above, font=\scriptsize, pos=0.62] {fails} (rej.west);
\draw[arr] (t2.east) -- (bnd.west);
\draw[arr] (bnd) -- (claims);
\end{tikzpicture}
\caption{From solver to proven bounds. Untrusted components are orange, and trusted
ones are blue. The solver returns a certificate, which consists of an admissible
potential $v$ and a stream function $\psi$ whose curl is an exactly
charge-conserving current. The checker never
runs the solver. A certificate that fails the specification check is rejected.
Otherwise, numerical verification returns bounds that hold for the stated model
whatever the solver did, and claims are judged against them. Tiles are judged by
their RMS field rather than pointwise. A solver of the wrong equations passes the first
test, but here the bounds refute its claims or leave them unsupported. A mesh study
compares a solver only with itself and makes neither test. Certificates at a few
conductivity ratios are blended into bounds that hold for every ratio in between
(Section~\ref{sec:ranges}).}
\label{fig:process}
\end{figure}
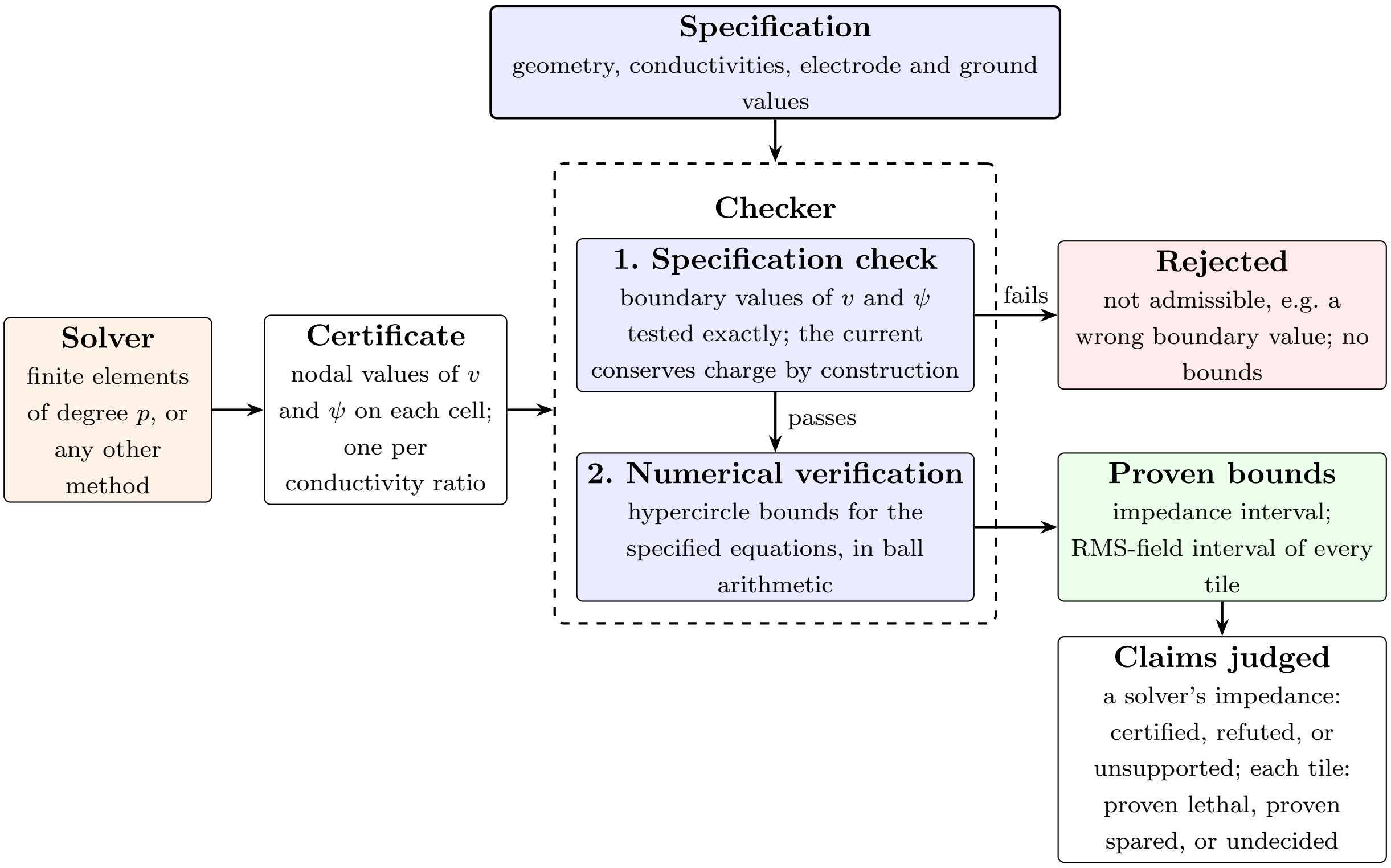

Figure~\ref{fig:process} shows how a solver's output becomes proven bounds.

\subsection{The Hypercircle}
Write $\norm{X}^2 = \int |X|^2\sigma^{-1}\dmu$, with $\norm{\cdot}_\omega$ the same
integral restricted to $\omega$. A potential $v\in H^1$ (with the weight $r$)
is \emph{admissible} if $v=u_D$ on $\Gamma_D$. A current $F\in H(\mathrm{div})$
is admissible if $\nabla\cdot F = 0$ and $F\cdot n=0$ on $\Gamma_N$.

\begin{theorem}[Prager--Synge \cite{pragersynge1947}]\label{thm:ps}
For admissible $v$ and $F$,
$\norm{\sigma\nabla v-\sigma\nabla u}^2+\norm{\sigma\nabla u-F}^2
=\norm{\sigma\nabla v-F}^2=:\eta^2$.
\end{theorem}
\begin{proof}
Let $w=v-u$, which vanishes on $\Gamma_D$. Then
$\int\nabla w\cdot(\sigma\nabla u-F)\dmu = 0$. The first term vanishes by the
weak form of the problem. In the second, integration by parts leaves a volume
term, which vanishes because $\nabla\cdot F=0$, and boundary terms, which vanish
because $w=0$ on $\Gamma_D$ and $F\cdot n=0$ on $\Gamma_N$. The identity
follows from Pythagoras' theorem.
\end{proof}
Hence, $\sigma\nabla u$ lies on the sphere with center $c=(\sigma\nabla v+F)/2$
and radius $\eta/2$. Restricting to a tile $\omega$ in one material and using
the triangle inequality gives $\norm{\sigma\nabla u}_\omega\in\norm{c}_\omega\pm\eta/2$
(the lower end clipped at zero), so
\begin{equation}\label{eq:tile}
E_\omega\in\bigl[(\norm{c}_\omega-\eta/2)_+,\ \norm{c}_\omega+\eta/2\bigr]\big/\sqrt{\sigma_\omega|\omega|}.
\end{equation}
The radius $\eta/2$ is global, so the bound is not localized, and its
half-width $\eta/(2\sqrt{\sigma_\omega|\omega|})$ is largest for the smallest
tiles, those on the axis.

For the conductance $G=\norm{\sigma\nabla u}^2/V^2$, the Dirichlet principle
gives $G\le\norm{\sigma\nabla v}^2/V^2$, and the sphere gives
$(\norm{c}-\eta/2)_+^2/V^2\le G\le(\norm{c}+\eta/2)^2/V^2$. Let $\Phi$ be the
current that $F$ carries off the driven electrode. In the bipolar case this is
the tip, and the current through the ring is $-\Phi$. In both configurations,
$\langle F,\sigma\nabla u\rangle=V\Phi$, so Cauchy--Schwarz gives Thomson's
bound $G\ge\Phi^2/\norm{F}^2$. For the bipolar catheter, $G$ is the
tip-to-ring conductance. The Dirichlet bound exceeds $G$ by
$\norm{\sigma\nabla(v-u)}^2/V^2$ and Thomson's falls short by a term of the same
order, so the impedance is certified to second order in $\eta$ (about $10^{-8}$
relative in Section~\ref{sec:results}) and a tile by \eqref{eq:tile} to first order.

\subsection{Certificate Format}
On each grid cell, $v$ is a tensor polynomial of degree $p$ in $(r,z)$. The
current is $F$, the curl of a Stokes stream function $\psi$, which is a tensor
polynomial of degree $p$ in $(r^2,z)$:
\begin{equation}
F=\Bigl(-\frac{1}{r}\partial_z\psi,\ \frac{1}{r}\partial_r\psi\Bigr).
\end{equation}
This $F$ is divergence-free by construction, and continuity of $\psi$ gives
continuity of the normal current across cells. The flux of $F$ through a wall
is $2\pi$ times the change of $\psi$ along it. So $F\cdot n=0$ on $\Gamma_N$
exactly when $\psi$ is constant along each connected insulating wall. The axis
belongs to such a wall with $\psi=0$. For the floating bipolar catheter, the axis, the outer boundary, and the shaft above the ring form a single wall with
$\psi=0$. The spacer between the electrodes keeps its own constant, which is
$\Phi/(2\pi)$. Regularity on the axis needs both $\psi=0$ there and the variable
$r^2$; with $r$ in its place, $F_z$ would behave like $1/r$. The certificate
consists of nodal values on each cell, equispaced in $r$ for $v$ and in $r^2$
for $\psi$. The solver may compute them in any way.

\subsection{The Checker}
The checker reads the specification and the nodal arrays, and never runs a
solver. It makes two tests. The specification check is steps 1 and 2 below,
and numerical verification steps 3 and 4:
\begin{enumerate}
\item It verifies the grid against the geometry.
\item It verifies admissibility exactly, namely the electrode, ring, and ground values of $v$ and the zero or constant value of $\psi$ along each insulating wall, at every
boundary node, including those inside cell edges (a degree-$p$ polynomial
constant at the $p+1$ nodes of an edge is constant on that edge).
\item It rebuilds each cell's polynomials by an exact change of basis.
\item It integrates every inner product exactly.
\end{enumerate}
On a cell $r_1\le r\le r_1+h_r$, with $r=r_1+h_r\xi$, each integrand is a polynomial
in $\xi$ plus a term $N(\xi)/(r_1+h_r\xi)$ from the $1/r$ in $F$. Near the axis, this
term is handled by exact polynomial division. On thin cells ($h_r\ll r_1$), where division would lose precision to cancellation, it is handled
by moments obtained from a downward recurrence with a rigorous starting
enclosure. All arithmetic is in 256-bit balls \cite{arb}, and every float in the
certificate is read as the exact binary number it is. The trusted base is the
specification (about 220 lines of Python), the checker and range checker (about
790 lines), the ball-arithmetic library python-flint, and NumPy, which is used
for the boundary masks and the exact equality tests.

\paragraph{Outcomes.} The specification check decides whether a certificate is
admissible. A run is \emph{rejected} if it is not, for example when a boundary
value differs from the specified one, and then no bounds follow. For an admissible
certificate, numerical verification always returns proven bounds for the stated
model, which include the discretization, solver, and rounding errors of the run.
The bounds come from the certificate alone. A solver may also report its own
value of a certified quantity, here the impedance; we call this value its claim
and judge it against the bounds. The claim is
\emph{refuted} if it lies outside the certified interval, and \emph{unsupported}
if the interval is too wide to prove it to a tolerance set by the user (1\% in
Section~\ref{sec:gci}). The last outcome
occurs for an inaccurate solver and for a solver of the wrong equations alike.
Otherwise, the claim is \emph{certified}, which means that it is correct for the
stated model to within that tolerance. Section~\ref{sec:gci} shows all three outcomes. Each tile is judged in
the same way, as proven lethal, proven spared, or undecided.

\subsection{A Vectorized Checker}\label{sec:fast}
The checker above works cell by cell in 256-bit arithmetic, which keeps it easy to
audit. A vectorized checker computes the same exact cell integrals on whole arrays
of cells in double-precision ball arithmetic \cite{rump1999,rump2010}. Each
quantity is an array of midpoints and an array of radii. Each operation adds a
bound on the rounding of its midpoint to the radius, and each radius, itself
computed in round-to-nearest, is inflated by $1+\gamma_k$, with
$\gamma_k=k\epsilon/(1-k\epsilon)$, the unit roundoff $\epsilon=2^{-53}$, and $k$ at least the number of roundings in its
computation. The bounds are rigorous under four assumptions:
\begin{itemize}
\item binary64 arithmetic with round-to-nearest;
\item no flush-to-zero of subnormal numbers, which the checker tests when it runs;
\item sums and matrix products formed by correctly rounded operations in some
order, possibly with fused multiply--add, but not by Strassen-type algorithms;
\item the binary64 values of the specification's parameters taken as the specified values.
\end{itemize}
The $1/r$ moments, once per grid column, and the final roots and divisions are
computed in Arb. The vectorized code adds about 600 lines to the trusted base.

On the lesion-map certificate of Section~\ref{sec:results}, it reproduces the
bounds of the 256-bit checker. Its $\eta$ agrees to $2.6\times10^{-6}$ relative,
its impedance interval differs by about $10^{-9}$\,\si{\ohm}, and every tile
interval overlaps the 256-bit one. It takes \SI{1.09}{s}, about a tenth of
our solver's time (Table~\ref{tab:cost}). Its cost grows linearly with the number of cells,
while the cost of the solve grows faster, so its share falls to 4.0\% at 6.3
million nodal values. A vectorized version of the range checker of
Section~\ref{sec:ranges} works in the same way. On the 17 certificates of
Section~\ref{sec:rangeresults}, it takes \SI{30}{s} instead of \SI{466}{s}, against \SI{154}{s}
for the solves, with $\eta_{\max}$ agreeing to $2.5\times10^{-6}$ relative and the same
lesion extents. All bounds reported below
come from the 256-bit checkers.

\begin{table}[t]
\centering
\caption{Cost of Checking Against Solving}\label{tab:cost}
\small
\begin{tabular}{lrrrrr}
\toprule
Spacing & Nodal values & Solve & Vectorized check & Share of solve & 256-bit check\\
\midrule
\SI{0.5}{mm} & 157\,360 & \SI{2.2}{s} & \SI{0.33}{s} & 15\% & \SI{5.6}{s}\\
\SI{0.25}{mm} & 506\,476 & \SI{9.5}{s} & \SI{1.09}{s} & 11\% & \SI{17.8}{s}\\
\SI{0.125}{mm} & 1\,749\,265 & \SI{52.8}{s} & \SI{3.58}{s} & 6.8\% & \SI{59.8}{s}\\
\SI{0.0625}{mm} & 6\,270\,646 & \SI{333}{s} & \SI{13.2}{s} & 4.0\% & --\\
\bottomrule
\end{tabular}
\par\vspace{2pt}\footnotesize\raggedright Monopolar electrode, degree 3, spacing in the lesion zone; the \SI{0.25}{mm} grid is the lesion-map certificate. ``Nodal values'' counts every node of the degree-3 node grid, including inactive ones. MacBook Pro (M5 Pro, 48 GB). Both checks are rigorous; --: 256-bit check not run.
\end{table}

\section{Bounds Over Parameter Ranges}\label{sec:ranges}

The electrodes are voltage-controlled, so multiplying both conductivities by
one factor leaves $u$ unchanged. Field quantities therefore depend only on
$\rho=\sigma_b/\sigma_t$. We keep $\sigma_t$ at its specified value and let
$\sigma_b=\rho\sigma_t$ range over an interval. Under current control, this
reduction would fail, because the field then scales with $1/\sigma$.

\begin{proposition}
Let $(v_k,F_k)$ and $(v_{k+1},F_{k+1})$ be certificates on the same grid, with
the same Dirichlet data, computed for ratios $\rho_k<\rho_{k+1}$. For every
$s\in[0,1]$, $v(s)=(1-s)v_k+s v_{k+1}$ and $F(s)=(1-s)F_k+sF_{k+1}$ are
admissible for every $\rho$.
\end{proposition}
Admissibility involves no conductivity, and both conditions are preserved by
convex combinations. Theorem~\ref{thm:ps} therefore holds at each $\rho$ with
any $s$. We choose $s=(\rho-\rho_k)/\Delta$ with $\Delta=\rho_{k+1}-\rho_k$, and
write $\eta(\rho)$ for the resulting radius. Let $\sigma_k=\sigma_t\rho_k$ be
the blood conductivity of certificate $k$ and $e_k=\sigma_k\nabla v_k-F_k$ its
own residual in blood. Since $\sigma(\rho)-\sigma_k=\sigma_t\Delta s$ and
$\sigma(\rho)-\sigma_{k+1}=-\sigma_t\Delta(1-s)$, the residual in blood is
\begin{equation}
\begin{aligned}
\sigma(\rho)\nabla v(s)-F(s)={}&(1-s)\,e_k+s\,e_{k+1}\\
&+\sigma_t\Delta\,s(1-s)\,(\nabla v_k-\nabla v_{k+1}),
\end{aligned}
\end{equation}
and in tissue the last term is absent. So $\eta(\rho)^2$ is a quadratic form in
$(1-s,s,\sigma_t\Delta s(1-s))$ of inner products among small fields, weighted
by $1/\sigma(\rho)$. In this form, ball evaluation over sub-intervals of $s$
avoids the cancellation that the naive expansion
$\int(\sigma|\nabla v|^2-2\nabla v\cdot F+|F|^2/\sigma)\dmu$ would suffer. The tile centers, which lie in tissue, are $c_k+s(c_{k+1}-c_k)$ with
$c_k=(\sigma_t\nabla v_k+F_k)/2$. The checker accumulates these inner products
once. It then evaluates \eqref{eq:tile} on 64 sub-intervals of $s$ per
segment; ball evaluation is inclusion-monotone, so the bounds hold for every
$\rho$ in the interval. The interpolation term shrinks as $\Delta^2$, so a
modest number of samples brings $\eta$ back to the level of a single
certificate. The approach is close in spirit to majorants for problems with
indeterminate data \cite{mali2014}, here with blended certificates and rigorous
evaluation. Threshold ranges follow directly, since a tile is lethal for every
threshold in $[E_1,E_2]$ if its lower bound reaches $E_2$. Catheter--tissue
gaps change the geometry and are certified one value at a time.

\section{Implementation and Verification}\label{sec:impl}

The solver is our own finite-element code, and it is untrusted. The checker
trusts no solver, so the bounds do not depend on which one produced the
certificate. We wrote our own because standard packages do not output the
stream-function current that a certificate needs; Section~\ref{sec:others}
certifies a potential computed by FEniCSx with the same checker and the same
result (Table~\ref{tab:fenics}). Our solver uses degree-$p$ Lagrange elements for
$v$ and fits $\psi$ by least squares so that $F$ approximates $\sigma\nabla v$. Where an electrode meets an insulator, the potential behaves
like $\sqrt{\text{distance}}$. The squared energy error of the first cell then
scales with its size; without grading, 99\% of $\eta^2$ lay in the two cells at the
electrode-shaft junction. Geometric grading down to \SI{1}{nm} at the junction and the tip edge
adds only a logarithmic number of grid lines. This junction singularity comes from the
change of boundary condition and remains for a rounded electrode; the sharp tip edge
adds a weaker one, which rounding would remove.
Together with $p=3$, the grading brings $\eta$ from $1.64\times10^{-3}$ (degree~1 with
weaker grading, $1.3\times10^5$ cells) to $1.3\times10^{-5}$ (degree~3,
$169\times332$ cells). Relative to $\norm{c}$, the final value is about
$1.3\times10^{-4}$.

The test suite checks the following:
\begin{itemize}
\item Certificates from the correct solver at different degrees, resolutions, and voltages give
overlapping bounds.
\item Perturbing a single electrode, ring, or ground value, or a stream-function
value on an insulating wall, by as little as $10^{-9}$ leads to rejection,
also at nodes inside cell edges.
\item A solver that drops the axisymmetric weight, or swaps the two
conductivities, still emits admissible certificates. Evaluated with the true
physics, the solver's own claims are refuted or unsupported.
\item With insulating blood, the certified impedance lies inside the window
given by Newman's value $1/(4\sigma_t a)$ for a disk on a half-space
\cite{newman1966}, corrected for the finite ground.
\item On a coarse problem, independent certificates at conductivity ratios not
used to build a range certificate lie inside the range bounds.
\item The sampled field of the solver lies inside the certified tile intervals.
\item The vectorized checkers agree with the 256-bit ones, also on noisy certificates.
\end{itemize}
The script \texttt{experiments/paper\_checks.py} reproduces the remaining checks
quoted below. On the full map, independent certificates at $\rho=2.2$ and
$\rho=3.3$, which were not used to build the range certificate, lie inside the
range bounds on all 768 tiles.

\section{Results}\label{sec:results}

\subsection{Focal Monopolar Electrode}
\begin{figure}[t]
\centering
\includegraphics[width=0.68\textwidth]{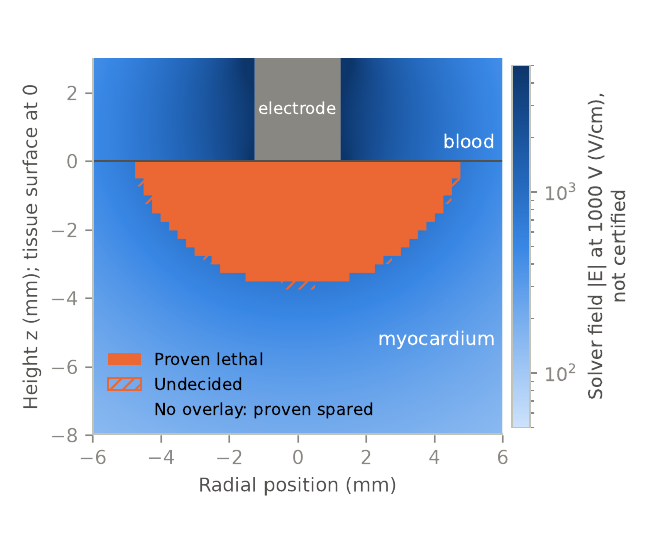}
\caption{Certified lesion of the focal monopolar electrode at \SI{1000}{V},
shown mirrored about the axis. Orange tiles are proven lethal
($E_\omega\ge\SI{700}{V/cm}$), hatched tiles are undecided, and all others are
proven spared. The background is the solver's field, which is not
certified.}\label{fig:map}
\end{figure}

\begin{figure}[t]
\centering
\includegraphics[width=0.8\textwidth]{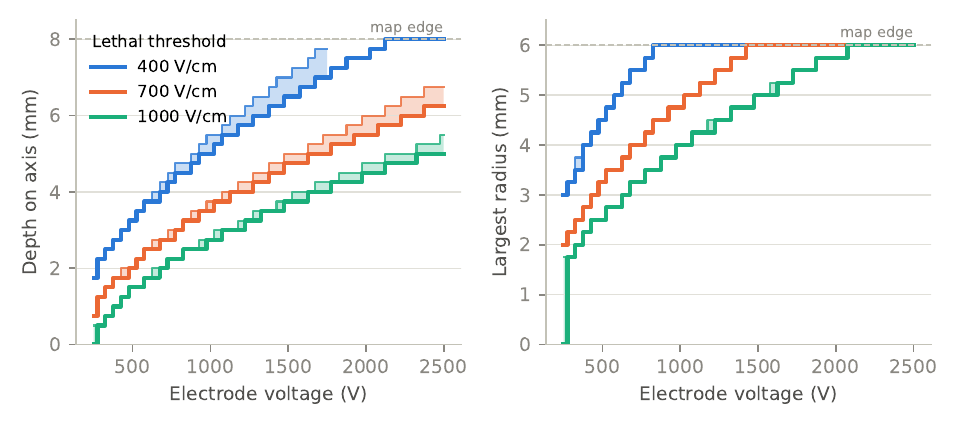}
\caption{Certified lesion depth and largest radius against voltage. Lines show
proven reached (for the depth, on the axis); bands extend to proven not beyond,
over every column and row of tiles. The map ends at
\SI{8}{mm} depth and \SI{6}{mm} radius.}\label{fig:extent}
\end{figure}

For the monopolar electrode ($L=\SI{3}{mm}$, $\sigma_t=\SI{0.25}{S/m}$,
$\sigma_b=\SI{0.7}{S/m}$, nominal ratio $\rho=2.8$, contact), the certified impedance is
\SIrange{93.950274}{93.950276}{\ohm}. For this idealized geometry, the
value indicates how tight the bounds can be rather than how accurate the model
is. On the $24\times32$ map of
\SI{0.25}{mm} tiles, the median relative width (width over midpoint) of the RMS-field intervals is
about 2\%. Fig.~\ref{fig:map} shows the certified lesion at \SI{1000}{V} and
\SI{700}{V/cm}. It is proven to reach \SI{3.50}{mm} on the axis and proven not
to extend beyond \SI{3.75}{mm} in any column; for the radius, reached and not beyond coincide
at \SI{4.75}{mm} at tile resolution. Only five
tiles on the threshold contour remain undecided. Fig.~\ref{fig:extent} gives
the certified extent against voltage for three thresholds. The map covers
\SI{6}{mm} of radius and \SI{8}{mm} of depth. At \SI{700}{V/cm} and from about \SI{1500}{V}, the
lesion can reach its edge, and those extents are reported as ``edge'' rather
than extrapolated. The ratio of the certified width (twice the radius) to the depth, about 2.6, is larger
than the value of about 2 that Castellv\'i and Ivorra report for conventional
isotropic threshold models of focal monopolar catheters \cite{castellvi2025};
part of the difference may come from the sharp electrode corner, which inflates
the radius at the surface.

\subsection{Conductivity and Threshold Uncertainty}\label{sec:rangeresults}
Blood conductivities of \SIrange{0.6}{0.75}{S/m} and myocardial conductivities
of \SIrange{0.2}{0.35}{S/m} give ratios within $[1.7,3.75]$. Seventeen certificates
cover this interval with $\eta_{\max}=1.52\times10^{-5}$. The median and worst
relative widths of the tile intervals are 8\% and 27\%, including the variation
over $\rho$. With five certificates, interpolation error raised
$\eta_{\max}$ to $1.4\times10^{-4}$ and widened the proven depth band to
\SIrange{3.0}{4.5}{mm}. This widening came from interpolation rather than from the conductivity. Point certificates at $\rho=1.7$, 2.2, 3.3, and 3.75 each prove a
depth between \SI{3.50}{mm} and \SI{3.75}{mm}. Table~\ref{tab:ranges} shows that, within this range,
the conductivity barely moves the lesion, while a $\pm\SI{100}{V/cm}$
uncertainty in the threshold widens the proven band considerably.

\begin{table}[t]
\centering
\caption{Certified Lesion of the Monopolar Electrode for the Nominal and for All Conductivity Ratios}\label{tab:ranges}
\small\setlength{\tabcolsep}{4pt}
\begin{tabular}{llcccc}
\toprule
Voltage & Threshold (V/cm) & Depth, nominal & Depth, all $\rho$ & Radius, nominal & Radius, all $\rho$\\
\midrule
\SI{500}{V} & 700 & 2.00--2.00 & 2.00--2.00 & 3.25--3.25 & 3.00--3.25\\
\SI{500}{V} & 600--800 & 1.75--2.50 & 1.75--2.50 & 3.00--3.50 & 2.75--3.50\\
\SI{1000}{V} & 700 & 3.50--3.75 & 3.50--3.75 & 4.75--4.75 & 4.75--5.00\\
\SI{1000}{V} & 600--800 & 3.25--4.00 & 3.25--4.25 & 4.50--5.25 & 4.25--5.25\\
\SI{1500}{V} & 700 & 4.75--5.00 & 4.50--5.00 & 6.00--edge & 5.75--edge\\
\SI{1500}{V} & 600--800 & 4.25--5.50 & 4.25--5.50 & 5.50--edge & 5.50--edge\\
\SI{2000}{V} & 700 & 5.50--6.00 & 5.50--6.00 & 6.00--edge & 6.00--edge\\
\SI{2000}{V} & 600--800 & 5.00--6.50 & 5.00--6.75 & 6.00--edge & 6.00--edge\\
\bottomrule
\end{tabular}
\par\vspace{2pt}\footnotesize\raggedright Entries are proven reached to proven not beyond (mm), for the nominal ratio $\rho=2.8$ and for every $\rho\in[1.7,3.75]$; ``edge'' means the lesion may reach the edge of the map.
\end{table}

\subsection{Comparison With a Grid Convergence Study}\label{sec:gci}
A practitioner would verify the monopolar solution by solving on
systematically refined grids and applying Richardson extrapolation with the
grid convergence index (GCI) \cite{roache1998,celik2008}. We did this with the
same solver at degree~2, on two families of four grids with refinement ratio 2
and spacing from \SI{1}{mm} to \SI{0.125}{mm} in the lesion zone. One family is
uniform near the electrode, as a typical user would mesh it; the other is graded
toward the electrode corners (the tip edge and the electrode-shaft junction) as
in Section~\ref{sec:impl}. The graded family scales the grading slope and the
far-field growth with the spacing but keeps its smallest spacing at \SI{1}{nm}.
Both families keep grid lines at the geometry and tile boundaries. The grids are
therefore not geometrically similar near the corners, and the observed orders
there should be read with this in mind. On every grid,
we recorded what the finite-element solution alone gives (the impedance from
the field energy and the tile RMS field from the potential gradient) and the
certified interval for the same quantity. The GCI uses the three finest grids
and a safety factor of 1.25. The convergence ratio $R$ classifies each triplet;
the GCI is valid only for monotone convergence, $0<R<1$ \cite{stern2001}. The
reference is a degree-3 certificate on the finest graded grid.
Table~\ref{tab:gci} summarizes the results for the impedance and two tiles at
\SI{1000}{V}, one at the electrode edge and one on the axis at the lesion boundary.

On the graded grids, both methods work. Every triplet converges at an observed
order of about 4, and every GCI band is consistent with the reference. For the
impedance, the certified interval is as narrow as the GCI band. Even here, the
extrapolated impedance, \SI{93.9502738}{\ohm}, lies just below the certified
reference interval; the difference of \SI{6e-7}{\ohm} is irrelevant
in practice but shows that extrapolation carries no guarantee. For the two
tiles, the certified intervals, about $\pm0.3\%$ and $\pm7\%$, are far wider
than the GCI bands (about $\pm10^{-5}\%$), because every tile bound uses the
global $\eta$. Where the grid is
good, a careful mesh study is sharper for local quantities, although it remains
an estimate.

The uniform grids behave differently. The impedance still converges, but at
an observed order of 0.19, far from the asymptotic range, so the GCI band is
$\pm1.3\%$; the certificate from the finest grid alone is about nine times
narrower and proven. At the tile touching the electrode edge, refinement makes
the change larger rather than smaller ($R=5.2$), so the triplet diverges and the
GCI is not valid. Applied regardless, it reports $\pm0.33\%$ around a value
0.6\% above the upper end of the reference interval, a band that excludes the
true value. The certified
interval on the same grid is wide but contains the reference, and its width
indicates that the grid is inadequate near the edge. On the axis
tile, the triplet also diverges, slowly ($R=1.14$), and the certified interval
is too wide to be useful.

A mesh study checks convergence rather than correctness. Because it compares a
solver only with itself, even an expert running it cannot tell whether the value it
converges to is the one the specification defines. We repeated it on the graded
grids with the two solvers of Section~\ref{sec:impl} that model the wrong
physics. Both converge cleanly, at an observed order of about 4, and the GCI
gives tight bands around their own impedances, \SI{136.806}{\ohm} with the
conductivities swapped and \SI{0.393}{\ohm} without the axisymmetric weight,
against a true \SI{93.950}{\ohm}, so a mesh study accepts both. Checked against
the specification on the finest grid, the claim of the solver without the axisymmetric weight lies outside
its certified interval, \SIrange{40.3}{145.4}{\ohm}, and is refuted. The solver with swapped conductivities
receives \SIrange{88.6}{220.9}{\ohm}, which contains its claim but cannot prove
it to 1\%, so the claim is unsupported; on the
same grid the correct solver's interval is $10^{7}$ times narrower, and its claim,
\SI{93.950270}{\ohm}, lies inside it and is certified. Code
verification with manufactured solutions or a review of the setup could also catch
such errors, but as separate work rather than as part of every run, and the manufactured source needs an independent statement of the physics.

The two methods are therefore complementary for conventional solvers. A mesh
study estimates the discretization error and is often sharp in the asymptotic
range. The certificate gives a guaranteed bracket for the specified problem,
which falsifies a GCI band disjoint from it, corroborates one it tightly
contains, and says nothing either way when it is wide. It holds on every grid
from a single solve. Its present weakness is that local intervals are wider
than a converged GCI band. The script \texttt{experiments/gci\_comparison.py} reproduces both comparisons.

\begin{table}[t]
\centering
\caption{Grid Convergence Index Versus Certified Intervals for the Monopolar Electrode}\label{tab:gci}
\small\setlength{\tabcolsep}{3pt}
\resizebox{\textwidth}{!}{%
\begin{tabular}{llcccc}
\toprule
Quantity & Grids & Finite element, finest grid & GCI: $R$, order, band & Certified, finest grid & Reference\\
\midrule
Impedance (\si{\ohm}) & Uniform & 93.8163 & 0.88, 0.19, 92.60--95.03 & 93.8163--94.0818 & \multirow{2}{*}{93.9502744--93.9502758}\\
 & Graded & 93.950270 & 0.06, 3.96, 93.950264--93.950275 & 93.950269--93.950280 & \\
Edge tile (V/cm) & Uniform & 5708 & 5.2, n/a, 5690--5727 & 3306--8027 & \multirow{2}{*}{5664.1--5674.7}\\
 & Graded & 5669.4 & 0.06, 4.00, 5669.4025--5669.4038 & 5654.7--5684.1 & \\
Axis tile (V/cm) & Uniform & 693.6 & 1.14, n/a, 691.6--695.6 & 0--8522 & \multirow{2}{*}{676.1--710.9}\\
 & Graded & 693.5 & 0.07, 3.91, 693.4809--693.4812 & 644.7--742.2 & \\
\bottomrule
\end{tabular}}
\par\vspace{2pt}\footnotesize\raggedright Degree~2, four grids per family with spacing \SIrange{1}{0.125}{mm}; tile RMS fields at \SI{1000}{V}. The GCI uses the three finest grids; $R>1$ means the triplet diverges, the GCI is not valid, and no order is given. Reference: a degree-3 certificate on the finest graded grid, hence tighter than the interval of Section~\ref{sec:results}. Edge tile: $1.25\le r\le\SI{1.5}{mm}$, $-0.25\le z\le\SI{0}{mm}$; axis tile: $0\le r\le\SI{0.25}{mm}$, $-3.75\le z\le\SI{-3.5}{mm}$.
\end{table}

\subsection{Bipolar Catheter and Comparison With Meckes et al.}
Meckes et al.\ modeled a two-electrode catheter with \SI{2}{mm} electrodes
\SI{4}{mm} apart. They used a lethal threshold of \SI{268}{V/cm} and compared
their predictions with in-vivo lesion depths \cite{meckes2022}. Their
conductivities are not given in the main text. Meckes et al.\ do
not state the catheter orientation, the electrode radius, or whether the spacing is
measured edge to edge. We assume a catheter
perpendicular to the tissue (axisymmetric), an electrode radius of
\SI{1.25}{mm}, a \SI{4}{mm} insulating spacer, a floating generator, and no temperature dependence. We replace the missing conductivities by a range. At contact, 17 certificates cover $\rho\in[1.7,3.75]$ with
$\eta_{\max}=1.7\times10^{-5}$. Gaps of 1--\SI{4}{mm} are certified at the
nominal ratio. With Thomson's bound, the tip-to-ring impedance at contact is
certified as \SI{172.6044}{\ohm} at the nominal ratio, tight to the last
digit shown. The median and worst relative widths of the range tile
intervals are 18\% and 68\%.

\begin{table}[t]
\centering
\caption{Certified Depth of the Bipolar Catheter Compared With Meckes et al.}\label{tab:meckes}
\small
\begin{tabular}{lccl}
\toprule
Voltage & Certified (mm) & Meckes model (mm) & In vivo (mm), as reported by Meckes et al.\\
\midrule
\SI{500}{V} & 2.75--3.00 & 2.28 & $2.25\pm0.85$ (1.35--3.14)\\
\SI{2200}{V} & 5.75--7.50 & 6.16 & $6.5\pm1.7$ (3.8--9.2)\\
\bottomrule
\end{tabular}
\par\vspace{2pt}\footnotesize\raggedright Bipolar catheter at contact, \SI{268}{V/cm}. Certified: proven reached to proven not beyond, for every $\rho\in[1.7,3.75]$. In vivo: mean $\pm$ SD as reported by Meckes et al., 95\% confidence interval in parentheses.
\end{table}

\begin{table}[t]
\centering
\caption{Voltage Needed for a Given Depth With a Blood Gap Under the Tip}\label{tab:gap}
\small
\begin{tabular}{lcc}
\toprule
Gap & \SI{1}{mm} deep & \SI{3}{mm} deep\\
\midrule
\SI{0}{mm} & 150 / 175 V & 475 / 550 V\\
\SI{1}{mm} & 225 / 275 V & 675 / 775 V\\
\SI{2}{mm} & 400 / 450 V & 975 / 1175 V\\
\SI{3}{mm} & 650 / 750 V & 1350 / 1750 V\\
\SI{4}{mm} & 950 / 1175 V & 1800 / ${>}2500$ V\\
\bottomrule
\end{tabular}
\par\vspace{2pt}\footnotesize\raggedright Nominal ratio; largest voltage proven insufficient / smallest voltage proven sufficient, on a \SI{25}{V} grid.
\end{table}

\begin{figure}[t]
\centering
\includegraphics[width=0.66\textwidth]{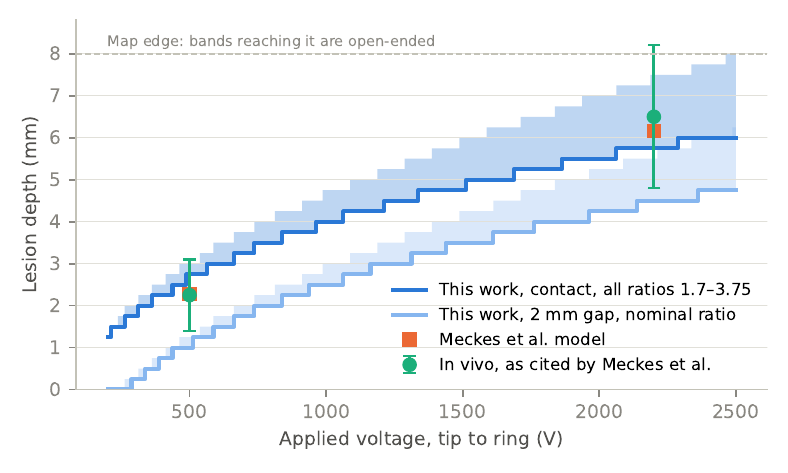}
\caption{Bipolar catheter: certified depth against voltage at contact (every
$\rho\in[1.7,3.75]$) and at a \SI{2}{mm} gap (nominal $\rho$). Lines show proven
reached and bands extend to proven not beyond; bands reaching the \SI{8}{mm} map
edge are open-ended. At high voltage, the threshold is met deep in tissue, where
the field decays slowly, so several tiles lie within their certified
uncertainty of the threshold and the band widens. The squares are Meckes et
al.'s model predictions and the circles the in-vivo depths they cite, with
$\pm$ one standard deviation.}\label{fig:meckes}
\end{figure}

Table~\ref{tab:meckes} and Fig.~\ref{fig:meckes} show that the certified
depths overlap the in-vivo data at both voltages. These data come from other
studies cited by Meckes et al., at \SI{500}{V} and \SI{2200}{V}, with
different catheters and waveforms, and their
intervals are wide. The overlap is therefore a plausibility check and does not
validate the model. For the gap, Meckes et al.\
report that 1--\SI{2}{mm} roughly doubles the voltage needed for \SI{3}{mm};
the strong effect of electrode-tissue proximity on PFA lesions has also been
studied in tissue \cite{howard2022}.
Table~\ref{tab:gap} agrees at \SI{2}{mm}, where the factor lies between
1.77 and 2.47, and gives a smaller factor at \SI{1}{mm}, between 1.23 and 1.63. At low
voltage, our lesions are deeper than in their model. We reach \SI{3}{mm} with at
most \SI{550}{V}, whereas Meckes et al.\ need \SI{700}{V}, and \SI{1}{mm} with
at most \SI{175}{V} against their \SI{300}{V}. At \SI{1200}{V} and a \SI{4}{mm}
gap, we certify \SIrange{1.0}{1.5}{mm} where they report a negligible lesion.

The certificates narrow down where this difference can come from. They prove
that it is not numerical error in our computation; a converged mesh study would
indicate the same without proof. They also exclude every constant conductivity
ratio in $[1.7,3.75]$. Outside that interval, the depth does depend on the ratio
(Table~\ref{tab:rho}). A lower ratio gives a shallower lesion. Within
constant-conductivity models,
$\rho=0.7$ reproduces two of the figures of Meckes et al. The certified depth at
\SI{500}{V} is \SIrange{2.25}{2.50}{mm}, which contains their \SI{2.28}{mm},
and \SI{700}{V} is proven sufficient for \SI{3}{mm}, their reported minimum.
This ratio means myocardium more conductive than blood, \SI{1.0}{S/m} at
$\sigma_b=\SI{0.7}{S/m}$, about three times the top of the
\SIrange{0.2}{0.35}{S/m} range, which is at the edge of plausibility. Their model
includes field- and temperature-dependent conductivity, but such a model raises
the conductivity most near the electrode, where the field is strongest, and in
sequential electroporation models this can enlarge rather than shrink the
electroporated region \cite{sel2005}. Whether their nonlinearity acts like a lower effective ratio
cannot be decided without their model. Even at $\rho=0.7$, the \SI{1}{mm}
threshold differs. We need at most \SI{200}{V} against their \SI{300}{V}.

\begin{table}[t]
\centering
\caption{Bipolar Catheter: Point Certificates Inside and Outside the Conductivity Range}\label{tab:rho}
\small\setlength{\tabcolsep}{3pt}
\begin{tabular}{lccc}
\toprule
& Depth at & \SI{3}{mm} & \SI{1}{mm}\\
$\rho=\sigma_b/\sigma_t$ & \SI{500}{V} (mm) & needs (V) & needs (V)\\
\midrule
0.7 & 2.25--2.50 & 650 / 700 & 175 / 200\\
1.0 & 2.50--2.50 & 575 / 650 & 150 / 175\\
2.8 & 2.75--3.00 & 475 / 550 & 150 / 175\\
7.0 & 3.00--3.00 & 425 / 500 & 125 / 175\\
10 & 3.00--3.25 & 425 / 500 & 125 / 175\\
Meckes et al. & 2.28 & 700 & 300\\
\bottomrule
\end{tabular}
\par\vspace{2pt}\footnotesize\raggedright Contact, \SI{268}{V/cm}; voltages as proven insufficient / proven sufficient.
\end{table}

Other candidates are further differences between the two models:
\begin{itemize}
\item Catheter orientation (a perpendicular tip drives current straight into
the wall; a catheter lying along it does not).
\item Electrode radius and spacing.
\item Boundary conditions.
\item Field- and temperature-dependent conductivity, as discussed above.
\item The lesion criterion (tile RMS here, pointwise there, a difference of
about one tile where the field decreases monotonically).
\item Numerical error in the reference model, which certificates of our own
model cannot exclude.
\end{itemize}
Deciding among these needs their exact model and, for field-dependent
conductivity and orientation, the extensions of Section~\ref{sec:discussion}.
A disagreement that remains after certification therefore lies in the physics or in
the other computation.

\section{Discussion}\label{sec:discussion}

\subsection{What a Certificate Establishes} Every bound is a theorem about the
stated model, that is, about the geometry, conductivities, and boundary
conditions as written in the specification. Provided the checker and the specification are correct,
the interval contains the stated model's true value whatever a solver returns,
an inadmissible certificate is rejected, and a claim that is wrong by more than the
tolerance is refuted or unsupported (Section~\ref{sec:cert}). Trust therefore rests on the checker and the
specification. The checker itself needs code verification, which the test suite
provides in part. The theorems used here are short. Formalizing them in a proof
assistant, together with a verified checker, would leave only the specification and a proof kernel to trust.

A tight certificate verifies the computed solution of that run. In the monopolar
example, $\eta$ is about $1.3\times10^{-4}$ of $\norm{c}$. The computed current
density is therefore proven to lie within 0.013\% of the exact solution of the
stated problem in the energy norm, although the exact solution is never known,
just as a two-sided bracket proves digits of $\sqrt2$. A verified solver would
give the same conclusion for this input. The certificate reaches it for a single
run, a far smaller task than verifying a solver for all inputs. A wide bound is
also informative, because by Theorem~\ref{thm:ps}, $\sigma\nabla v$ or $F$ is then at least
$\eta/\sqrt2$ from the exact current density in the energy norm. A number computed from them may still be accurate, and such a claim is then
unsupported rather than refuted.

A wrong specification, with wrong units, geometry, or conductivity (use error in
V\&V~40 terms), is certified as faithfully as a right one. A wrong model is a
separate failure, addressed by validation and in part by certifying over
parameter ranges, which replaces ``for our guessed conductivity'' by ``for every
conductivity in this range''. The lethal threshold is itself contested. Petras
et al.\ find that no single threshold fits both depth and width, with roughly
\SIrange{270}{500}{V/cm} for depth and \SIrange{790}{1000}{V/cm} for width
\cite{petras2025}.

\subsection{Use in Credibility Assessment} Certificates provide rigorous
calculation verification for every run in the context of use, rather than for a
representative case. For patient-specific clinical tools, the highest example
level of discretization-error evidence proposed by Galappaththige et al.\ is a
convergence analysis performed automatically on each new patient, with the error
estimated and used in processing the results \cite{galappaththige2022}; a
certificate replaces that estimate by a proven bound. The inner envelope supports
conclusions about efficacy at the target, at tile resolution, whereas conclusions
about the safety of collateral structures need pointwise bounds.
Certificates address the numerical part of credibility, on which the rest
depends. A comparison with data is informative only once the numerical error is
known to be small. Validation, the applicability of the validation evidence, and
uncertainty quantification remain separate tasks, and they carry most of the
weight in medical device submissions
\cite{pathmanathan2017,viceconti2021,galappaththige2022}. Range
certificates also contribute to uncertainty quantification.

With the vectorized checkers (Section~\ref{sec:fast}), a check can accompany every run,
including a range certificate, whose 17 ratios take \SI{154}{s} of solves and \SI{30}{s}
of checking.

\subsection{Other Solvers and Learned Surrogates}\label{sec:others} In principle, any solver that
outputs fields can participate. A conventional finite-element solution provides
an admissible potential, and a flux reconstruction provides the current, for
instance by local equilibration \cite{braess2009,ernvohralik2015} or by the
least-squares stream-function fit our solver uses. A finite-volume solution
provides the current, and a reconstruction provides the potential.

As a check of independence, the potential of the monopolar lesion-map problem
was also computed by FEniCSx \cite{baratta2023}, with its own assembly,
quadrature, and linear solver. It used the same grid and the same degree-3
tensor elements, and the stream-function fit, used as an untrusted adapter,
supplied the current. Table~\ref{tab:fenics} shows that the unchanged checker
gives the same bounds as for our solver. The two potentials differ by at most
\SI{7e-9}{V}, and FEniCSx's own impedance, computed from the energy of its potential, equals
the certified lower bound, as it must, since that bound is the energy of the same
potential. This tests an independent implementation on the same
discretization. Certificates for FEniCSx on its own meshes are given in
\cite{dovelos2026b}.

\begin{table}[t]
\centering
\caption{The Same Check on Two Independent Solvers}\label{tab:fenics}
\small
\begin{tabular}{lccc}
\toprule
Solver of the potential & Certified impedance (\si{\ohm}) & $\eta$ & Tile width, median / worst\\
\midrule
Our solver & 93.950274296--93.950275789 & $1.3\times10^{-5}$ & 1.9\% / 18\%\\
FEniCSx 0.11 & 93.950274296--93.950275789 & $1.3\times10^{-5}$ & 1.9\% / 18\%\\
\bottomrule
\end{tabular}
\par\vspace{2pt}\footnotesize\raggedright Monopolar lesion map, contact, nominal ratio; same grid and degree-3 tensor elements; current from the stream-function fit in both rows. Bounds rounded outward.
\end{table}

For PFA, neural networks already predict the ablated area, validated
against experimental data from the literature \cite{crusi2025}. Their error
comes from approximation and training, so Richardson extrapolation and the GCI
do not apply, and test-set and conformal guarantees hold in probability over an
input distribution rather than for a given output. A surrogate output that is
made admissible can be certified like any other, and the bound remains valid
outside the training distribution. Bounds of this functional a posteriori type
have been proposed for neural solutions \cite{fanaskov2024}; the independent
checker, the rigorous arithmetic, and the written specification are what this
paper adds. An output can be made admissible by post-processing, which without
Galerkin orthogonality needs a global correction \cite{papez2018}, or by
construction, with exact wall constants for the stream function and a potential
lifted to the Dirichlet data; \cite{dovelos2026b} demonstrates the second.

\subsection{Limitations and Extensions} The model is axisymmetric, linear,
isotropic, and quasi-static, with constant conductivities. It has no frequency
dispersion, no temperature dependence, and no field dependence; the last would
also break the linearity in $V$ that lets one certificate serve all voltages.
Electrodes are voltage-controlled flat-tipped cylinders with sharp corners, and
the corner touching the tissue inflates the lesion radius at the surface. A
rounded electrode would remove it but needs a checker for curved boundaries. The
monopolar return is a grounded box \SI{50}{mm} away rather than a skin patch.

Real lesions depend on more than this model contains. The lethal threshold
depends on the waveform and on whether lesions are read acutely or chronically
\cite{kotnik2019,verma2021}. The tissue has finite wall thickness, so the
clinical question is transmurality, and it contains fat, scar, and trabeculation.
Electrodes indent the tissue under contact force, repeated high-voltage trains
heat tissue and can form gas at the electrodes, and clinical multielectrode
catheters produce three-dimensional fields. The perpendicular tip and
tip-to-ring configurations here are stylized test cases for the method.

The method has limits of its own. Lesion criteria use tile RMS fields. Gaps are
certified at sampled values only. The comparison with Meckes et al.\ rests on
our assumptions about their geometry. Bands widen with depth because every tile
bound uses the global $\eta$, so reducing $\eta$ would tighten them and smaller
tiles would not. Pointwise bounds could be added, since in each homogeneous
region $|\nabla u|^2$ is subharmonic and its value at a point is at most its
mean over any ball around the point that lies in that region.

The orientation question of Section~\ref{sec:results} needs a three-dimensional
model. Nothing in the hypercircle bound or in the checker's logic depends on axisymmetry;
only the certificate format does. On a tetrahedral mesh,
the potential is a continuous piecewise polynomial with the electrode and ground
values. The current is best supplied as the curl of a N\'ed\'elec vector
potential, the three-dimensional analog of the stream function used here,
because a Raviart--Thomas field with floating-point coefficients is not exactly
divergence-free. Alternatively, a small residual divergence can be bounded by a
term involving a Friedrichs constant \cite{repin2008}. The checker then
integrates products of polynomials over tetrahedra in closed form, and the range
method of Section~\ref{sec:ranges} carries over unchanged. A polygonal mesh does not
represent a curved electrode exactly, so the specification would state the meshed
geometry, or the geometric error would need a separate bound. Electrode edges become
singular lines that need anisotropic grading, and meshes of $10^5$ to $10^6$
elements need a vectorized checker.

Field-dependent conductivity, the other candidate in
Section~\ref{sec:results}, needs no third dimension. With a conductivity that
increases with $|\nabla u|$, the problem still minimizes a convex energy, and a
convex-duality analog of Theorem~\ref{thm:ps} gives certificates of the same
kind \cite{repin2008}. For semilinear problems, equilibrated fluxes already
certify existence and outputs \cite{ishizaka2026}.

\section{Conclusion}

When simulations run without an expert checking each result, the evidence of
accuracy has to come with each result. We have shown that each run can be verified by its own
certificate. The solver supplies an admissible potential and current, and a
small checker, independent of it, turns them into guaranteed bounds against a
written specification of the physics. Its vectorized version checks a single
certificate in about a tenth of our solver's time, and a range certificate in a
fifth of the time of its solves. For PFA, the bounds certify lesion maps at the
\SI{0.25}{mm} tile resolution near the threshold contour at moderate voltages,
with bands that widen with depth. They extend to whole ranges of uncertain
tissue conductivity, they remain valid on grids where a standard mesh study
does not converge, and they rule out our own numerical error as the cause of a
disagreement with another model. For PFA device modeling, this means that each
predicted lesion, for a given catheter, voltage, and range of tissue
conductivity, comes with a proven bound on its numerical error, so that a comparison with
tissue data tests the model alone. For conventional solvers, mesh studies
and certificates are complementary. Learned surrogates
have no grid to refine, so a guarantee for a given output needs an a posteriori
bound of this kind. The method is limited today to linear,
axisymmetric conduction. Certificates for other solvers and learned surrogates
are given in \cite{dovelos2026b}; localized bounds and three-dimensional geometry
are the next steps.

\section*{Code Availability}
The solver, checker, tests, and the scripts that reproduce every number in this
paper are available at \url{https://github.com/kostisdov/certsim} (tag
\texttt{v1.1-preprint}) under the Apache License 2.0.


\begin{thebibliography}{99}
\bibitem{meckes2022} D.~Meckes, M.~Emami, I.~Fong, D.~H.~Lau, and P.~Sanders, ``Pulsed-field ablation: Computational modeling of electric fields for lesion depth analysis,'' \emph{Heart Rhythm O2}, vol.~3, no.~4, pp.~433--440, 2022, doi: 10.1016/j.hroo.2022.05.009.
\bibitem{petras2025} A.~Petras \emph{et al.}, ``Is a single lethal electric field threshold sufficient to characterize the lesion size in computational modeling of cardiac pulsed-field ablation?'' \emph{Heart Rhythm O2}, vol.~6, no.~5, pp.~671--677, 2025, doi: 10.1016/j.hroo.2025.02.014.
\bibitem{anic2026} A.~Ani\'{c}, A.~Ivorra, B.~Koop, N.~J.~Kirkpatrick, and J.~Koruth, ``Pulsed field ablation: From concepts to technology,'' \emph{Europace}, vol.~28, no.~Suppl.~2, pp.~ii18--ii31, 2026, doi: 10.1093/europace/euag205.
\bibitem{asmevv40} American Society of Mechanical Engineers, \emph{Assessing Credibility of Computational Modeling Through Verification and Validation: Application to Medical Devices}, ASME Standard V\&V 40-2018, New York, NY, USA, 2018.
\bibitem{fda2023} U.S. Food and Drug Administration, ``Assessing the credibility of computational modeling and simulation in medical device submissions,'' Guidance for Industry and FDA Staff, Docket FDA-2021-D-0980, Nov. 2023.
\bibitem{roache1998} P.~J.~Roache, \emph{Verification and Validation in Computational Science and Engineering}. Albuquerque, NM, USA: Hermosa, 1998.
\bibitem{oberkampf2010} W.~L.~Oberkampf and C.~J.~Roy, \emph{Verification and Validation in Scientific Computing}. Cambridge, U.K.: Cambridge Univ. Press, 2010, doi: 10.1017/CBO9780511760396.
\bibitem{lu2021} L.~Lu, P.~Jin, G.~Pang, Z.~Zhang, and G.~E.~Karniadakis, ``Learning nonlinear operators via DeepONet based on the universal approximation theorem of operators,'' \emph{Nat. Mach. Intell.}, vol.~3, no.~3, pp.~218--229, 2021, doi: 10.1038/s42256-021-00302-5.
\bibitem{dovelos2026b} K.~Dovelos, ``Verifying the outputs of PDE solvers: From finite elements to neural surrogates,'' arXiv:2611.31211, 2026.
\bibitem{necula1997} G.~C.~Necula, ``Proof-carrying code,'' in \emph{Proc. 24th ACM SIGPLAN-SIGACT Symp. Princ. Program. Lang. (POPL)}, Paris, France, 1997, pp.~106--119, doi: 10.1145/263699.263712.
\bibitem{mcconnell2011} R.~M.~McConnell, K.~Mehlhorn, S.~N\"aher, and P.~Schweitzer, ``Certifying algorithms,'' \emph{Comput. Sci. Rev.}, vol.~5, no.~2, pp.~119--161, 2011, doi: 10.1016/j.cosrev.2010.09.009.
\bibitem{pragersynge1947} W.~Prager and J.~L.~Synge, ``Approximations in elasticity based on the concept of function space,'' \emph{Quart. Appl. Math.}, vol.~5, no.~3, pp.~241--269, 1947, doi: 10.1090/qam/25902.
\bibitem{synge1957} J.~L.~Synge, \emph{The Hypercircle in Mathematical Physics}. Cambridge, U.K.: Cambridge Univ. Press, 1957.
\bibitem{polya1951} G.~P\'olya and G.~Szeg\H{o}, \emph{Isoperimetric Inequalities in Mathematical Physics}. Princeton, NJ, USA: Princeton Univ. Press, 1951, doi: 10.1515/9781400882663.
\bibitem{ladeveze1983} P.~Ladev\`eze and D.~Leguillon, ``Error estimate procedure in the finite element method and applications,'' \emph{SIAM J. Numer. Anal.}, vol.~20, no.~3, pp.~485--509, 1983, doi: 10.1137/0720033.
\bibitem{braess2009} D.~Braess, V.~Pillwein, and J.~Sch\"oberl, ``Equilibrated residual error estimates are $p$-robust,'' \emph{Comput. Methods Appl. Mech. Eng.}, vol.~198, no.~13--14, pp.~1189--1197, 2009, doi: 10.1016/j.cma.2008.12.010.
\bibitem{ernvohralik2015} A.~Ern and M.~Vohral\'ik, ``Polynomial-degree-robust a posteriori estimates in a unified setting for conforming, nonconforming, discontinuous Galerkin, and mixed discretizations,'' \emph{SIAM J. Numer. Anal.}, vol.~53, no.~2, pp.~1058--1081, 2015, doi: 10.1137/130950100.
\bibitem{repin2008} S.~Repin, \emph{A Posteriori Estimates for Partial Differential Equations}. Berlin, Germany: De Gruyter, 2008, doi: 10.1515/9783110203042.
\bibitem{mali2014} O.~Mali, P.~Neittaanm\"aki, and S.~Repin, \emph{Accuracy Verification Methods: Theory and Algorithms}. Dordrecht, The Netherlands: Springer, 2014, doi: 10.1007/978-94-007-7581-7.
\bibitem{liuoishi2013} X.~Liu and S.~Oishi, ``Verified eigenvalue evaluation for the Laplacian over polygonal domains of arbitrary shape,'' \emph{SIAM J. Numer. Anal.}, vol.~51, no.~3, pp.~1634--1654, 2013, doi: 10.1137/120878446.
\bibitem{liu2021stokes} X.~Liu, M.~T.~Nakao, C.~You, and S.~Oishi, ``Explicit a posteriori and a priori error estimation for the finite element solution of Stokes equations,'' \emph{Jpn. J. Ind. Appl. Math.}, vol.~38, no.~2, pp.~545--559, 2021, doi: 10.1007/s13160-020-00449-5.
\bibitem{nakao2019} M.~T.~Nakao, M.~Plum, and Y.~Watanabe, \emph{Numerical Verification Methods and Computer-Assisted Proofs for Partial Differential Equations}. Singapore: Springer, 2019, doi: 10.1007/978-981-13-7669-6.
\bibitem{rozza2008} G.~Rozza, D.~B.~P.~Huynh, and A.~T.~Patera, ``Reduced basis approximation and a posteriori error estimation for affinely parametrized elliptic coercive partial differential equations,'' \emph{Arch. Comput. Methods Eng.}, vol.~15, no.~3, pp.~229--275, 2008, doi: 10.1007/s11831-008-9019-9.
\bibitem{fanaskov2024} V.~Fanaskov, A.~Rudikov, and I.~Oseledets, ``Neural functional a posteriori error estimates,'' arXiv:2402.05585, 2024.
\bibitem{eigel2016} M.~Eigel and C.~Merdon, ``Local equilibration error estimators for guaranteed error control in adaptive stochastic higher-order Galerkin finite element methods,'' \emph{SIAM/ASA J. Uncertain. Quantif.}, vol.~4, no.~1, pp.~1372--1397, 2016, doi: 10.1137/15M102188X.
\bibitem{papez2018} J.~Pape\v{z}, Z.~Strako\v{s}, and M.~Vohral\'ik, ``Estimating and localizing the algebraic and total numerical errors using flux reconstructions,'' \emph{Numer. Math.}, vol.~138, no.~3, pp.~681--721, 2018, doi: 10.1007/s00211-017-0915-5.
\bibitem{ishizaka2026} H.~Ishizaka, ``Low-order CR--RT equilibrated-flux certification for semilinear problems on anisotropic meshes,'' arXiv:2602.01636, 2026.
\bibitem{cheung2017} K.~K.~H.~Cheung, A.~Gleixner, and D.~E.~Steffy, ``Verifying integer programming results,'' in \emph{Integer Programming and Combinatorial Optimization (IPCO)}, Lecture Notes in Computer Science. Cham, Switzerland: Springer, 2017, pp.~148--160, doi: 10.1007/978-3-319-59250-3\_13.
\bibitem{arb} F.~Johansson, ``Arb: Efficient arbitrary-precision midpoint-radius interval arithmetic,'' \emph{IEEE Trans. Comput.}, vol.~66, no.~8, pp.~1281--1292, 2017, doi: 10.1109/TC.2017.2690633.
\bibitem{rump1999} S.~M.~Rump, ``Fast and parallel interval arithmetic,'' \emph{BIT Numer. Math.}, vol.~39, no.~3, pp.~534--554, 1999, doi: 10.1023/A:1022374804152.
\bibitem{rump2010} S.~M.~Rump, ``Verification methods: Rigorous results using floating-point arithmetic,'' \emph{Acta Numer.}, vol.~19, pp.~287--449, 2010, doi: 10.1017/S096249291000005X.
\bibitem{newman1966} J.~Newman, ``Resistance for flow of current to a disk,'' \emph{J. Electrochem. Soc.}, vol.~113, no.~5, p.~501, 1966, doi: 10.1149/1.2424003.
\bibitem{castellvi2025} Q.~Castellv\'i and A.~Ivorra, ``Computational multiscale modeling of pulsed field ablation considering conductivity and damage anisotropy reveals deep lesion morphologies,'' \emph{Int. J. Numer. Methods Biomed. Eng.}, vol.~41, no.~8, Art. no.~e70077, 2025, doi: 10.1002/cnm.70077.
\bibitem{celik2008} I.~B.~Celik, U.~Ghia, P.~J.~Roache, C.~J.~Freitas, H.~Coleman, and P.~E.~Raad, ``Procedure for estimation and reporting of uncertainty due to discretization in CFD applications,'' \emph{J. Fluids Eng.}, vol.~130, no.~7, Art.~no.~078001, 2008, doi: 10.1115/1.2960953.
\bibitem{stern2001} F.~Stern, R.~V.~Wilson, H.~W.~Coleman, and E.~G.~Paterson, ``Comprehensive approach to verification and validation of CFD simulations---Part~1: Methodology and procedures,'' \emph{J. Fluids Eng.}, vol.~123, no.~4, pp.~793--802, 2001, doi: 10.1115/1.1412235.
\bibitem{howard2022} B.~Howard \emph{et al.}, ``Effects of electrode-tissue proximity on cardiac lesion formation using pulsed field ablation,'' \emph{Circ. Arrhythm. Electrophysiol.}, vol.~15, no.~10, Art. no.~e011110, 2022, doi: 10.1161/CIRCEP.122.011110.
\bibitem{sel2005} D.~\v{S}el, D.~Cukjati, D.~Batiuskaite, T.~Slivnik, L.~M.~Mir, and D.~Miklav\v{c}i\v{c}, ``Sequential finite element model of tissue electropermeabilization,'' \emph{IEEE Trans. Biomed. Eng.}, vol.~52, no.~5, pp.~816--827, 2005, doi: 10.1109/TBME.2005.845212.
\bibitem{galappaththige2022} S.~Galappaththige, R.~A.~Gray, C.~Mendonca Costa, S.~Niederer, and P.~Pathmanathan, ``Credibility assessment of patient-specific computational modeling using patient-specific cardiac modeling as an exemplar,'' \emph{PLoS Comput. Biol.}, vol.~18, no.~10, Art. no.~e1010541, 2022, doi: 10.1371/journal.pcbi.1010541.
\bibitem{pathmanathan2017} P.~Pathmanathan, R.~A.~Gray, V.~J.~Romero, and T.~M.~Morrison, ``Applicability analysis of validation evidence for biomedical computational models,'' \emph{J. Verif. Valid. Uncertainty Quantif.}, vol.~2, no.~2, Art. no.~021005, 2017, doi: 10.1115/1.4037671.
\bibitem{viceconti2021} M.~Viceconti, F.~Pappalardo, B.~Rodriguez, M.~Horner, J.~Bischoff, and F.~Musuamba Tshinanu, ``In silico trials: Verification, validation and uncertainty quantification of predictive models used in the regulatory evaluation of biomedical products,'' \emph{Methods}, vol.~185, pp.~120--127, 2021, doi: 10.1016/j.ymeth.2020.01.011.
\bibitem{baratta2023} I.~A.~Baratta, J.~P.~Dean, J.~S.~Dokken, M.~Habera, J.~S.~Hale, C.~N.~Richardson, M.~E.~Rognes, M.~W.~Scroggs, N.~Sime, and G.~N.~Wells, ``DOLFINx: The next generation FEniCS problem solving environment,'' Zenodo, 2023, doi: 10.5281/zenodo.10447666.
\bibitem{crusi2025} R.~Crusi, N.~Colistra, F.~Camera, G.~Monti, M.~S.~Zappatore, C.~Merla, and L.~Tarricone, ``Cardiac pulsed-field ablation: Deep learning solutions for multi-parameter predictions,'' \emph{IEEE J. Electromagn. RF Microw. Med. Biol.}, vol.~9, no.~4, pp.~445--454, 2025, doi: 10.1109/JERM.2025.3577268.
\bibitem{kotnik2019} T.~Kotnik, L.~Rems, M.~Tarek, and D.~Miklav\v{c}i\v{c}, ``Membrane electroporation and electropermeabilization: Mechanisms and models,'' \emph{Annu. Rev. Biophys.}, vol.~48, no.~1, pp.~63--91, 2019, doi: 10.1146/annurev-biophys-052118-115451.
\bibitem{verma2021} A.~Verma, S.~J.~Asirvatham, T.~Deneke, Q.~Castellvi, and R.~E.~Neal~II, ``Primer on pulsed electrical field ablation: Understanding the benefits and limitations,'' \emph{Circ. Arrhythm. Electrophysiol.}, vol.~14, no.~9, 2021, doi: 10.1161/CIRCEP.121.010086.
\end{thebibliography}
\end{document}